\documentclass[conference]{IEEEtran}
\IEEEoverridecommandlockouts
\usepackage{cite}
\usepackage{amsmath,amssymb,amsfonts}
\usepackage{algorithmic}
\usepackage{graphicx}
\usepackage{textcomp}
\usepackage{xcolor}
\usepackage{booktabs}
\usepackage{amsmath}
\usepackage{soul}
\usepackage{tabularx}
\usepackage{caption}
\usepackage{algorithm}
\usepackage{comment}
\usepackage{xurl}
\def\BibTeX{{\rm B\kern-.05em{\sc i\kern-.025em b}\kern-.08em
    T\kern-.1667em\lower.7ex\hbox{E}\kern-.125emX}}

\newcommand{\reiot}{{$\mathrm{VEXA_{IoT}}$}}

\begin{document}

\title{\reiot: Autonomous IoT Vulnerability EXploitation using AI Agents \\
%{\footnotesize \textsuperscript{*}Note: Sub-titles are not captured in Xplore and
%should not be used}
%\thanks{Identify applicable funding agency here. If none, delete this.}
% Automated AI Agents driven 
}

%Autonomous IoT Vulnerability Exploitation using AI Agents
%Autonomous Exploitation of IoT Vulnerabilities using AI Agents
%Autonomous Exploitation of IoT Vulnerabilities using AI Agents
\iffalse
\author{\IEEEauthorblockN{Katherine Swinea}
\IEEEauthorblockA{\textit{Department of Computer Science} \\
\textit{Tennessee Tech University}\\
Cookeville, TN, USA \\
keswinea42@tntech.edu}
\and
\IEEEauthorblockN{Kshitiz Aryal}
\IEEEauthorblockA{\textit{School of Interdisciplinary Informatics} \\
\textit{University of Nebraska Omaha}\\
Omaha, NE, USA \\
karyal@nebraska.edu}
\and
\IEEEauthorblockN{Lopamudra Praharaj}
\IEEEauthorblockA{\textit{dept. name of organization (of Aff.)} \\
\textit{name of organization (of Aff.)}\\
City, Country \\
email address or ORCID}
\and
\IEEEauthorblockN{Maanak Gupta}
\IEEEauthorblockA{\textit{Department of Computer Science} \\
\textit{Tennessee Tech University}\\
Cookeville, TN, USA \\
mgupta@tntech.edu}
}
\fi

\author{
\IEEEauthorblockN{
Katherine Swinea\IEEEauthorrefmark{1},
Kshitiz Aryal\IEEEauthorrefmark{2}, Lopamudra Praharaj\IEEEauthorrefmark{3},
Maanak Gupta\IEEEauthorrefmark{1}}
\IEEEauthorblockA{\IEEEauthorrefmark{1}Department of Computer Science,
Tennessee Tech University, TN, USA.}
\IEEEauthorblockA{\IEEEauthorrefmark{2}School of Interdisciplinary Informatics, University of Nebraska Omaha, NE, USA.}\IEEEauthorblockA{\IEEEauthorrefmark{3}Dept. of Mathematics \&
Computer Science, University of North Carolina at Pembroke, NC, USA.}
Corresponding e-mail: mgupta@tntech.edu}
\maketitle

\begin{abstract}
Internet of Things (IoT) systems are inherently vulnerable due to constrained hardware, outdated firmware, and insecure default configurations, creating a need for scalable and adaptive security testing approaches. While recent adoptions of Large Language Model (LLM) agents have demonstrated promise in penetration testing and Capture-the-Flag (CTF) environments, their application to IoT specific vulnerabilities remains unexplored. This paper presents an autonomous multi-agent framework, referred to as \textbf{V}ulnerability \textbf{EX}ploitation using
\textbf{AI} Agents (\reiot), for vulnerability discovery and exploitation in IoT environments using LLM-based reasoning and offensive security tools. The framework combines a vulnerability detection agent and an attack execution agent to perform reconnaissance, plan attack sequences, and execute exploits against vulnerable IoT services. The system is evaluated in IoTGoat and Metasploitable environments across ten attack scenarios mapped to OWASP IoT vulnerabilities. Experimental results show attack success rate of up to 100\% with low token overhead and average execution times under two minutes for most attacks. Across 260 attack executions, \reiot~achieves a 95.0\% overall success rate, including 94.5\% success in IoTGoat and 96.7\% success in Metasploitable2. These results demonstrate the potential for LLM-driven agents to automate IoT vulnerability assessment and offensive security workflows in controlled environments.

% This has led to several projects studying vulnerabilities, exploits, and defenses. In recent years, artificial intelligence (AI) agents have been used for capture the flags (CTF) and penetration testing purposes. Using the purposefully vulnerable IoTGoat environment from OWASP, a framework is created to use large language model (LLM) agents to exploit these vulnerabilities and launch attacks. Results showed high success rates for 10 attacks with efficient token usage. The prompt given to the LLM uses information gathered from tools to determine the plan of attack for one agent. Another agent uses relevant information to execute an attack with the available tools. The agents use API calls to ChatGPT for the LLM reasoning. This allows the attacks to be run separately for a more sequential execution of the plan produced by the agent or in parallel, excluding any attacks that require information from another, to optimize the speed of all the attacks being completed. 

% \Kshitiz{Abstract is not clear, we need to motivate what we're doing and clearly explain our idea on a high level. Need to be rewritten at the end}
\end{abstract}


\section{Introduction} 

Internet of Things (IoT) devices have been largely integrated into modern society with the widespread adoption of smart devices in homes, healthcare, manufacturing, and industrial environments. The number of connected IoT devices continues to grow rapidly, reaching 18.5 billion devices in 2024 and increasing by 14\% in 2025\cite{sinha2025number}. The connected IoT devices are projected to exceed 39 billion by 2030. While this rapid growth has improved automation and connectivity, it has also expanded the attack surface across a large number of interconnected and often vulnerable devices\cite{Fortinet}. 

% Connected IoT devices were at 18.5 billion in 2024 and have increased by 14\% in 2025 showing that these devices are only becoming more prevalent \cite{b1}. It is also expected for connected devices to grow to 39 billion by 2030. The integration of these IoT devices into more networks can create security issues due to them being large interconnected networks of vulnerable devices \cite{b2}.

IoT devices have many known vulnerabilities that are difficult to address due to the constrained hardware limitations, limited computational capabilities, and infrequent firmware updates. Many devices often lack strong encryption methods as secure algorithms can be computationally intensive and resource-heavy \cite{8688434}. This can lead to problems such as unauthorized access to stored information or eavesdropping on communications that include sensitive information like credentials. Additionally, many IoT devices rely on hardcoded or default credentials that are publicly known or easily extracted from firmware images \cite{stanislav2015hacking}. These weaknesses can allow attackers to gain unauthorized access to the device with minimal effort. Even when vulnerabilities are identified, patching them remains difficult because update mechanisms themselves may lack integrity verification or authentication safeguards, enabling malicious firmware modifications or unauthorized updates \cite{BASNIGHT201376}. Together, these examples show how IoT devices can be easily compromised and that the issues are often intertwined making them persistent across IoT ecosystems. 

%\Kshitiz{Explain Few examples of IoT vulnerabilies here. How are these vulnerabilities in IoT devices exploited by attackers and offensive researchers traditionally (Reference to the existing works to motivate ours)? What are the challenges with current methods? We need to motivate the need and novelty of our work at this point. and How using agents help us overcome those challenges }

% These vulnerabilities are tested in an attempt to understand attack patterns and find defenses for them. Testbeds to simulate IoT devices or house them physically are used to test the effectiveness of different tools and determine the biggest issues. These often incorporate penetration testing tools that develop a modular way to run reconnaissance tools and attacks based on these tools \cite{b6}. These traditional methods require a lot of time and effort to test the vulnerabilities and run a successful attack from researchers. It is also difficult to scale these frameworks to larger systems or implement it in different settings due to the heterogeneous nature of IoT devices and the manual modifications needed to the modules and rule-based systems. This has caused a shift towards more adaptive methods using machine learning (ML). However, most of the solutions that have been presented are detection and classification methods rather than methods to perform and assess the attacks \cite{b6}. This leaves a need for a system that can adapt to different environments and decrease the time and effort needed to find and test vulnerabilities.

% \lopa{Penetration testing frameworks???} 

To better understand these vulnerabilities and develop effective defenses, researchers commonly use IoT testbeds and intentionally vulnerable environments to simulate real-world attack scenarios. Existing vulnerability scanning frameworks typically combine reconnaissance tools, exploit databases, and attack script in modular or rule-based workflows \cite{anand2020iot}. While these approaches are effective for targeted testing, they often require significant manual effort from security professionals to identify vulnerabilities, configure attacks, and adapt workflows to different environment. The challenge becomes more severe if you are to scale these frameworks to larger systems or implement it in different settings due to the heterogeneous nature of IoT devices as they frequently depend on static rules or environment-specific configurations \cite{Fortinet}. Although machine learning (ML) techniques have been introduced into IoT security research, most existing approaches focus on intrusion detection \cite{anomaly}, anomaly detection \cite{mobile}, and classification \cite{HASAN2019100059} of threats rather than autonomous vulnerability assessment and exploitation \cite{survey}. \textit{This creates a need for adaptive systems capable of autonomously identifying, testing, and exploiting vulnerabilities across diverse IoT environments.}

%\lopa{citations??} was after sentence that ends with "static rules or environment-specific configurations" Since the rules and envirments limitations were already brought up and cited in a previous sentence, I cited the article that talks about IoT security struggles.

%\lopa{Need to explain why we need the agentic framework here? may be with an example}

Recent advances in artificial intelligence (AI), particularly Large language models (LLMs) have demonstrated strong reasoning and decision-making capabilities across cybersecurity tasks \cite{inproceedings}. Agentic AI systems built on LLMs have been explored for penetration testing \cite{299699}, vulnerability analysis \cite{Xu2024AutoAttackerAL}, and Capture-the-Flag (CTF) environments \cite{unknown}. These systems have shown that general purpose LLMs can perform reconnaissance, generate attack strategies, and adapt to environmental feedback during offensive security tasks. However, much of the existing work focuses on solving predefined CTF challenges or executing highly specific attack objectives \cite{abramovich2025enigmainteractivetoolssubstantially}. While these studies demonstrate the potential of LLM-driven adversarial agents, they often lack flexibility for broader vulnerability-driven testing in real-world environments. Further, limited attention has been given to IoT specific implementations despite the unique security challenges and attack surfaces associated with the IoT devices.

% This shows the potential for agents using generic models that are not tailored to cybersecurity purposes. The agents were used for exploration and exploitation of vulnerabilities in red team scenarios. The focus has also been on how many attacks or CTF objectives an agent can complete \cite{b9}. While this shows how well agents can perform at specific tasks under specific circumstances, it does not provide much generality or flexibility in the approach. This brings attention to the need for a different approach for adversarial agents that achieve general goals over specific tasks in particular environment.This brings attention to the need for a different approach for adversarial agents that achieve general goals over specific tasks in particular environment.
%\Kshitiz{We need to elaborate here with other literatures that has been used for automated offensive/defensive security research using agents and their impact}

To address this gap, this work presents, Vulnerability EXploitation using AI
Agents (\reiot), an autonomous multi-agent framework for vulnerability discovery and exploitation in IoT environments. The framework combines a vulnerability detection agent with an attack execution agent to perform reconnaissance, analyze vulnerabilities, plan attack sequences, and execute exploits using common offensive security tools and practices. The system was evaluated using IoTGoat \cite{IoTGoat} and Metasploitable \cite{Rapid7} environments to demonstrate its ability to adapt to multiple vulnerable IoT-like systems and perform attacks mapped to OWASP IoT vulnerabilities \cite{OWASP}. Both implementations were able to run several attacks in the environment and most attacks had a 100\% success rate over 20 trials. %\Gupta{add a statement on the effectiveness of the approach/results} 
These results suggest that \reiot~can support automated penetration testing workflows in controlled vulnerable environments. The key contributions of this work are as follows:

% Clearly, \reiot~could be used to automate penetration testing in these environments. 

% Despite the interest in adversarial agents and the existence of particularly vulnerable domains like IoT, there has not been much focus on domain specific implementations. This leaves a variety of domain specific attacks untested by adversarial agents which are often able to go through different attacks faster than a human user and for longer periods of time. IoT devices are known to have a significant amount of issues and vulnerabilities that take time and effort to fully discover and exploit the way attackers would. This leaves a gap for using AI agents as adversaries to test IoT vulnerabilities and exploits. In order to address this, the contributions of this work are as follows:  

\begin{itemize}
      \item Designed a multi-agent framework for autonomous IoT vulnerability assessment and exploitation using LLM-based reasoning and offensive security tools.
      \item Developed a vulnerability detection agent that performs reconnaissance, identifies exposed services, and protocols, and analyzes known vulnerabilities using tools such as \texttt{nmap} and \texttt{searchsploit}.
      \item Developed an attack execution agent capable of selecting and executing attack-specific tools and scripts based on contextual vulnerability information and attack dependencies.
      % that determines how to executed an attack against the device based on the analysis from the vulnerability detection agent.}
      \item Designed prompt structures and reasoning workflows that enable coordinated attack planning, adaptive execution, and retry handling in response to failures or environmental feedback.  
      % that use role-playing techniques and presented vulnerability that consistently produced plans and instructions for attacks from the agents.}
      \item Evaluated the framework in IoTGoat and Metasploitable environments across multiple attack scenarios, demonstrating autonomous exploitation of vulnerabilities mapped to OWASP IoT security threats.
      % Tested these agents in an IoT environment to see real-time impacts and changes to the device as the agents launched attacks.}
\end{itemize}

%\Kshitiz{Contributions needs to be more specific, like developed agents for vulnerability scanning looking into these system components. ....}

The remaining sections of this paper are structured as follows. Section \ref{sec:background} covers related work on penetration testing agents and manual uses of IoTGoat. Section \ref{sec:framework} outlines the framework and methodology used to develop the environment, prompts, and agents. Section \ref{sec:implementation} presents and discuss the results from the attacks run by the agents. %Section V is a discussion based on the results and broader impacts of the results. 
Section \ref{sec:conc} concludes the paper and discusses future directions.

\section{Related Works}
\label{sec:background}
The use of LLMs for offensive cybersecurity tasks has gained significant attention in recent years \cite{threatgpt,lazer2026survey}, particularly in penetration testing and Capture-the-Flag (CTF) environments. Early research demonstrated that general-purpose LLMs could assist with vulnerability discovery, exploit generation, and attack planning through carefully designed prompts and reasoning workflows. Works like PentestGPT \cite{299699} and AutoAttacker \cite{Xu2024AutoAttackerAL} have used LLMs to generate penetration testing plans and have an autonomous system for attack execution respectively. However, these systems primarily focus on structured penetration testing workflows and do not specifically address IoT-oriented vulnerability exploitation.
% they have lacked the autonomy needed for an agent that to make adversarial decisions. 

Other research has investigated the ability of LLM agents to autonomously solve cybersecurity challenges. In \cite{unknown}, researchers demonstrated the potential for using a general-purpose LLM agents to solve CTF problems using ReAct-based prompting strategies combined with internal planning, adaptive retries, and environmental feedback, achieving success rates above 90\% in some scenarios. Works like AutoPentester \cite{11354838} take this a step further by implementing multiple agents and establishing a workflow between them to strategize, generate commands, and verification with the possibility to adjust commands. This pentesting agent also measured success based on completion of CTF problems and was able to achieve 100\% success for some cases. These systems showed that agentic workflows can improve offensive automation by enabling iterative reasoning and adaptive behavior during attack execution.

Despite these advances, most existing LLM-based offensive security research has focused on CTF environments or narrowly scoped penetration testing tasks. These environments typically involve predefined objectives, static challenge structures, and limited environmental variability. As a result, relatively little work has explored the use of autonomous adversarial agents in IoT-specific environments, where vulnerabilities, protocols, firmware configurations, and attack surfaces are significantly more heterogeneous and dynamic. 

% Unlike these systems, \reiot focuses on IoT-specific vulnerability workflows and evaluates attack execution across IoTGoat and a secondary vulnerable environment.

IoT security has been an active area of research due to the widespread adoption of vulnerable embedded devices and the growing number of attacks targeting the IoT ecosystems. The introduction of the OWASP IoT Top 10 \cite{OWASP} and environments such as IoTGoat \cite{IoTGoat} provided researchers and educators with intentionally vulnerable platforms for studying common IoT security issues. However, the existing practices on these IoT environments typically focus on manual use where users explore the system and find vulnerabilities and exploit themselves \cite{iot_secure}. Systems for automated adversarial IoT focus on more specific smart-home architectures and the specific OWASP vulnerabilities\cite{fi14100276}. This was a rule-based system based on the results of the attempted exploits. It did not have the flexibility of an agent that can adjust plans based on environmental feedback. It also only focused on finding and categorizing vulnerabilities without any ways to exploit these vulnerabilities. 

\begin{figure*}[h]
    \centering
    \includegraphics[width=\textwidth]{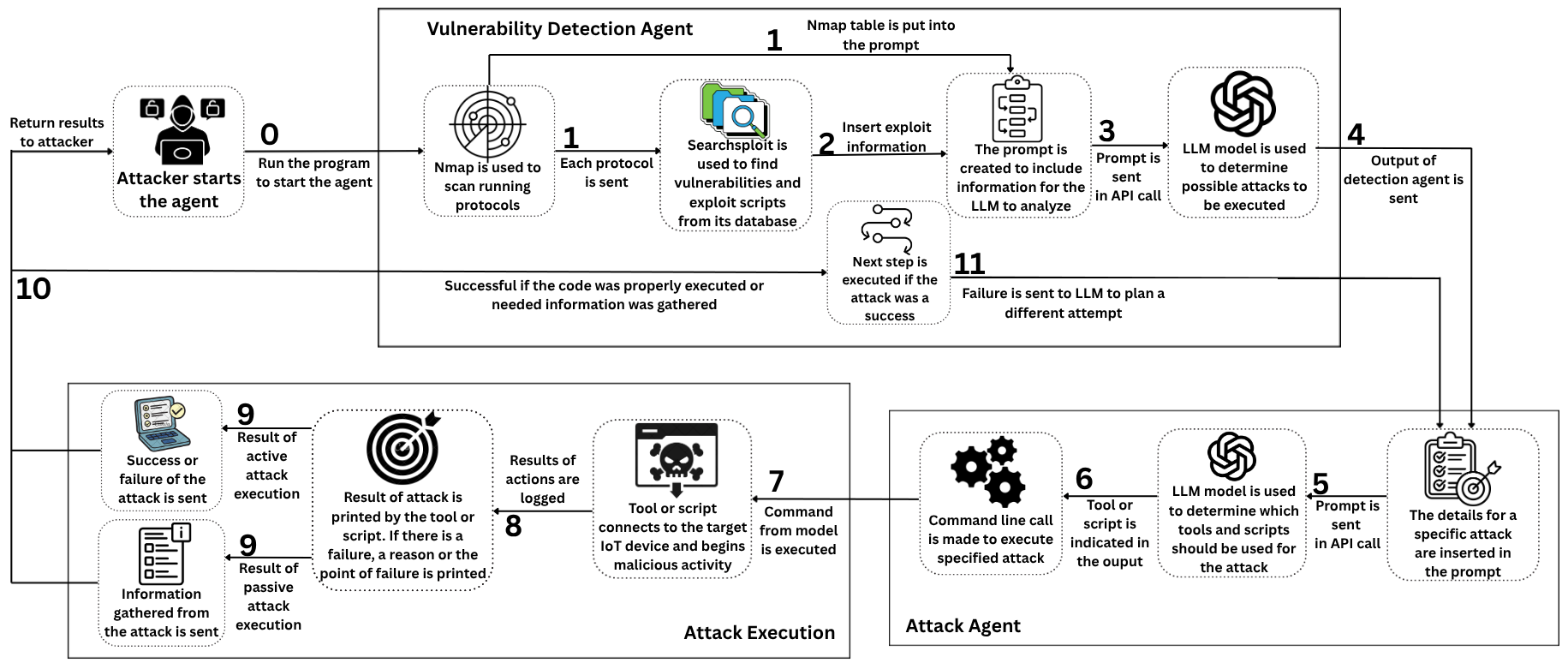}
    %\vspace{-18mm}
    \caption{Proposed \reiot~Framework for Vulnerability Scanning and Attack Execution}
    \label{fig:agentic}
\end{figure*}

%\hl{Font sizes of texts in figure needs to be increased}

\section{Proposed Framework and Methodology}
\label{sec:framework}
Our proposed \reiot~framework is designed to autonomously perform reconnaissance, vulnerability analysis, attack planning, and exploit execution in the IoT environment. \reiot~follows a multi-agent architecture consisting of two coordinated AI agents: a \textit{vulnerability detection agent} and an \textit{attack execution agent}. The vulnerability detection agent performs reconnaissance on the target environment, identifies exposed services and protocols, and analyzes potential vulnerabilities using publicly available exploit intelligence in \texttt{searchsploit's} Exploit Database \cite{searchsploit}. The attack execution agent then uses this information to determine appropriate attack methods, select offensive security tools, and execute exploits against the target device. It performs both, passive attacks that silently observe and gather information, and active attacks that execute malicious actions against the target. The \reiot~framework targets the OWASP IoT Top 10 vulnerabilities \cite{OWASP}: weak, guessable, or hardcoded passwords, insecure network services, insecure ecosystem interfaces, lack of secure update mechanisms, use of insecure or outdated components, insufficient privacy protection, insecure data transfer and storage, lack of device management, insecure default settings, and lack of physical hardening. Among these, only insecure data transfer is taken advantage of with a passive MitM attack. The remaining vulnerabilities are exploited with active attacks. %\Gupta{no mention of OWASP Top 10? - we need to discuss which attacks are done? active/passive} %Figure \ref{fig:agentic} presents the overall architecture of \reiot. 

% It accomplishes this with \textbf{two agents}. The \textbf{first agent} is a \textbf{vulnerability detection agent}. This agent focuses on finding the open ports and running protocols that can be used for vulnerability exploitation. Tools are used to determine known vulnerabilities for these protocols and the LLM model determines what exploits can be used. The attacks that should be executed are then passed to the \textbf{second agent, attack agent}. This agent determines what attack it will run and which tools should be used to execute it. The attack agent will give a command that is used to run the tool or script that will execute the attack. The output of the attack is then captured to ensure success before moving on to the next attack.

%\lopa{First begin with the border scope of the framework. Instead of chatgpt 5.0,the model should be more generic here}

\subsection{\reiot~Architecture}
The \reiot~architecture as shown in Figure \ref{fig:agentic} separates vulnerability analysis and exploit execution into two independent but coordinated workflows. This separation allows the system to maintain modularity while allowing adaptive attack planning based on the environmental feedback. %\Gupta{We should reference the steps in figure in discussion}

The vulnerability detection agent begins by scanning the target system using \texttt{nmap} \cite{nmap} to identify open ports, exposed services, and active network protocols, shown as step 1 in Figure \ref{fig:agentic}. In step 2, the results of the scan are processed using \texttt{searchsploit} which maps detected services and software versions to known CVEs \cite{searchsploit} and publicly available exploit scripts. Step 3 places the collected vulnerability information into the prompt provided to the LLM (ChatGPT 5.1 thinking), which analyzes the discovered services and determines an appropriate sequence of attacks based on identified vulnerabilities and attack dependencies. %\Gupta{which LLM we are using?}

% We use ChatGPT 5.1 thinking. I can move this information back up here, it was moved to the implementation section to keep the framework more generic. That is also why OWASP isn't mentioned here either. It can al so be moved back if that would be better. 

Once the attack plan is generated, the selected exploit and vulnerability information from the \texttt{searchsploit} lookup is passed to the attack execution agent as seen in step 4. This agent determines the specific tools or scripts required for the attacks in step 5 and generates commands necessary for execution in step 6. Step 7 executes the generated commands directly on the target system using offensive security tools such as \texttt{bettercap} \cite{bettercap}, and exploit scripts retrieved through \texttt{searchsploit} \cite{searchsploit}. The results of the attack execution are returned to the agent in step 8 including command outputs. The failure messages and attack confirmations are gathered in step 9 and returned to the vulnerability detection agent for validation in step 10 and orchestration of subsequent attacks in step 11.

The vulnerability detection agent continues to orchestrate the planned attacks, monitoring attack outcomes and determining whether the attack succeeded, failed or requires additional retries. Some attacks have dependencies on the information passive attacks collect. The detection agent confirms this information, gathered before moving on to the next attack. Active attacks instead confirm the attack succeeded and did not have any error codes or failure messages. This can help handle and possibly avoid an overall failure by adjusting the attack approach. The detection agent sends the failure and how the attack was executed to the attack agent. Attack agent will then attempt a different approach based on the reason for the failure.

\reiot is designed to operate autonomously once execution begins. The vulnerability detection agent orchestrates the workflow by calling the attack execution agent as a subprocess and parsing the resulting outputs to determine attack status. The decision to continue with the attack sequence or retry the attack in case of failure is handled in the detection agent. While the framework is capable of fully autonomous execution, all generated commands, attack plans, and execution outputs are logged and displayed to the user, enabling optional human oversight or intervention during runtime.

% These outputs are also printed in the terminal.Since each step and it's output is printed, it also allows for a user to take over by stopping the execution of the process. Otherwise, the agent will run until all the attacks are completed. 

\subsection{Prompt}
\begin{figure}[!t]
    \centering
    \includegraphics[width=\linewidth]{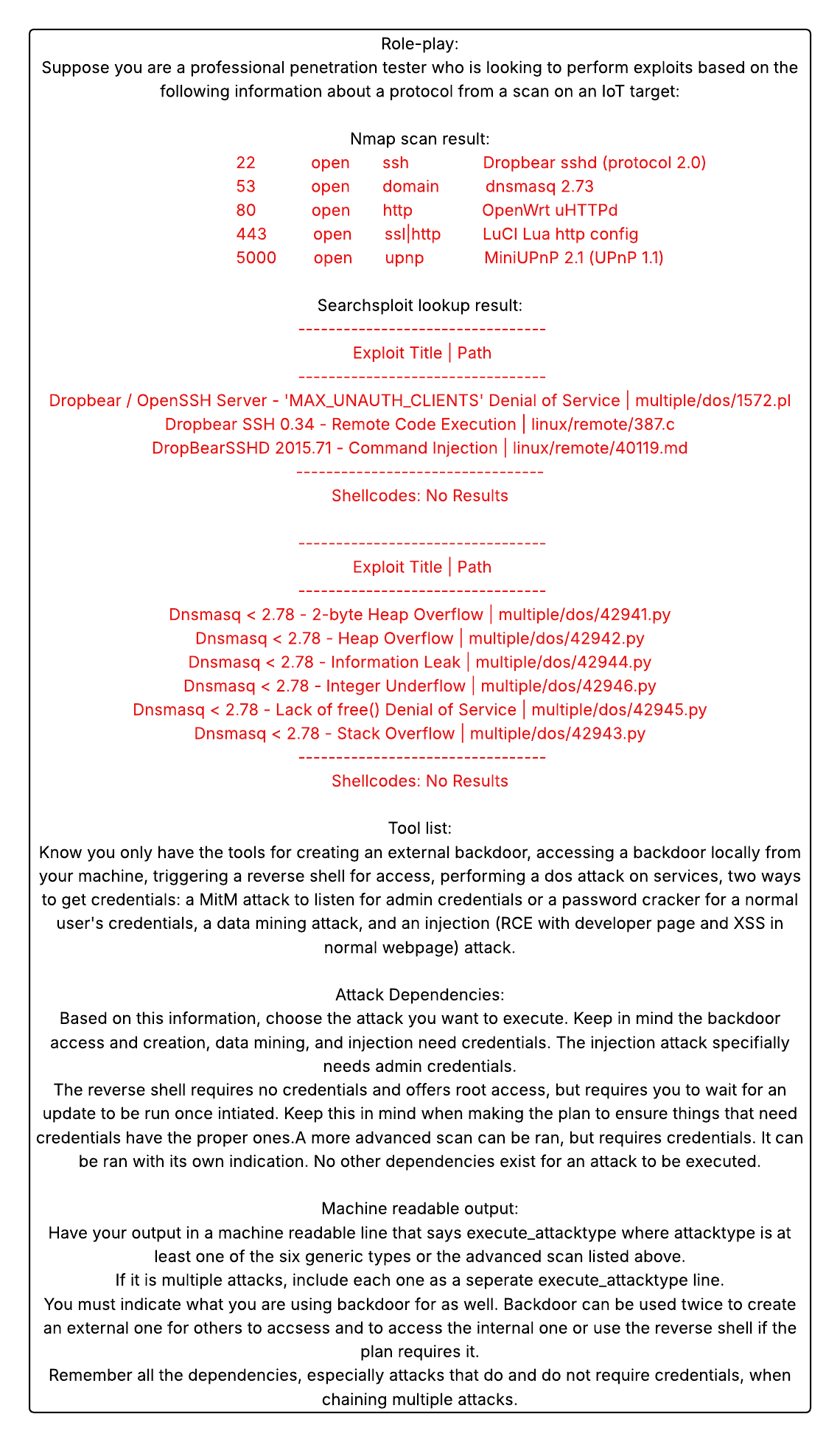}
    \caption{Example vulnerability detection prompt containing reconnaissance results, exploit candidates, tool capabilities, and dependency instructions.}
    \label{fig:Vulnerability_Det_Prompt}
\end{figure}

\reiot~uses prompt-based reasoning to guide both vulnerability analysis and attack execution. Two separate prompts are used: one for vulnerability detection agent and one for the attack execution agent.
Results from the tools and script execution are communicated to the LLM through the prompt. 
Both prompts use role-based task framing to improve consistency during controlled penetration-testing experiments. The prompts constrain the model to the isolated testbed environment, available tools, and user-defined objectives.
The vulnerability detection prompt includes the \texttt{nmap} scan results, \texttt{searchsploit} vulnerability information, available tools in the attacker environment, and optional user-defined goals. The tools and dependencies of the target machine are also outlined in the prompt to ensure the model plans attacks that can actually be executed. This prompt instructs the model to analyze detected services, identify exploitable vulnerabilities, determine attack dependencies, and generate an ordered attack sequence. Figure \ref{fig:Vulnerability_Det_Prompt} shows the template that is used for the vulnerability detection agent with an example of the \texttt{nmap} scan and \texttt{searchsploit} lookup results inserted into it.

% \caption{Prompt used by Vulnerability Detection Agent}

The prompt starts with specifying the role as a penetration tester and defining the environment that is being used. After that, the results from the \texttt{nmap} scan and \texttt{searchsploit} lookup are inserted as a variable. These inputs work as a ground truth for the vulnerabilities to ensure the agent is consistent when planning the attacks that should be used and the core reasoning for why those attacks were chosen. Rather than explicitly describing offensive tools as malicious utilities, the prompt describes the functional capabilities of each available tool. This approach reduced refusal behavior from the LLM while still allowing the model to reason about which tools were appropriate for a given attack objective. These functional descriptions of the tools help the detection agent determine which tools need to be used to achieve the user's goals. These goals are input by the user when calling the agent and stored in the scenario variable. If no goals are provided, the agent will run all the attacks. The prompt additionally includes instructions regarding attack dependencies and execution ordering to ensure prerequisite information, such as credentials or elevated privileges, is collected before dependent attacks are executed. 
A machine readable section of the output is specified in the execute\_attack format to easily parse the output for each attack.

%\Gupta{Incorrect reference of 4 and 5?}

\begin{figure}[!t]
    \centering
    \includegraphics[width=\linewidth]{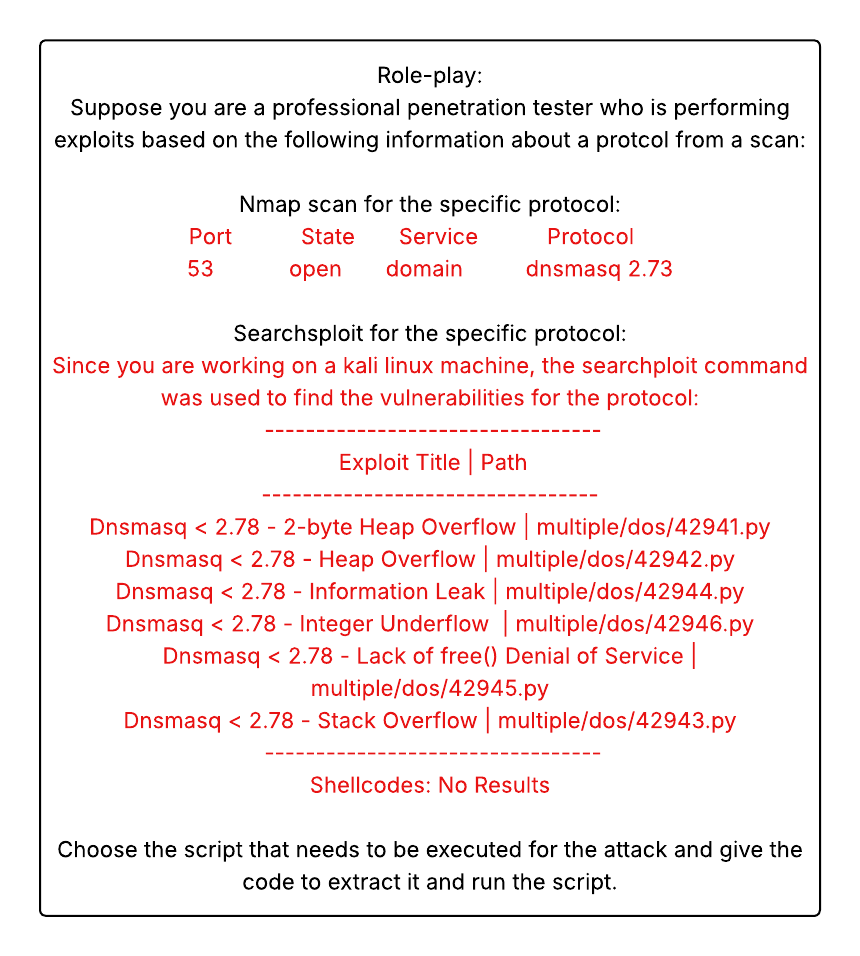}
    \caption{Example attack execution prompt for the DNS DoS scenario, including target service information and candidate exploit scripts.}
    \label{fig:attack_prompt}
\end{figure}

%\caption{Prompt for Attack Execution Agent}

The attack execution agent uses the exploit and vulnerability information from \texttt{searchsploit} and instructions for the attack's execution from the detection agent's output. Relevant detection information includes the port for the targeted protocol and the analysis of the vulnerability from \texttt{searchsploit} tool. The same generic prompt is used for all 10 attacks but the attack specific instructions can vary greatly. Figure \ref{fig:attack_prompt} shows the generic attack prompt with the DNS DoS information filled in. The protocol port scan information from \texttt{nmap} and the variety of options from the \texttt{searchsploit} lookup are included to specify the attack for the agent. The prompt also specifies that a command should be given to run the script that is chosen. The attack agent prompt also specifies the role but modifies it slightly to focus on executing the found exploits for a specific vulnerability rather than looking at all the possible vulnerabilities. The inputs from the detection agent guide the attack agent to ensure the correct attack is chosen. It chooses the tool based on the described functions and matches that to the indicated attack.

\subsection{Vulnerability Detection Agent}

\begin{figure}[!ht]
    \centering
    \includegraphics[width=\linewidth]{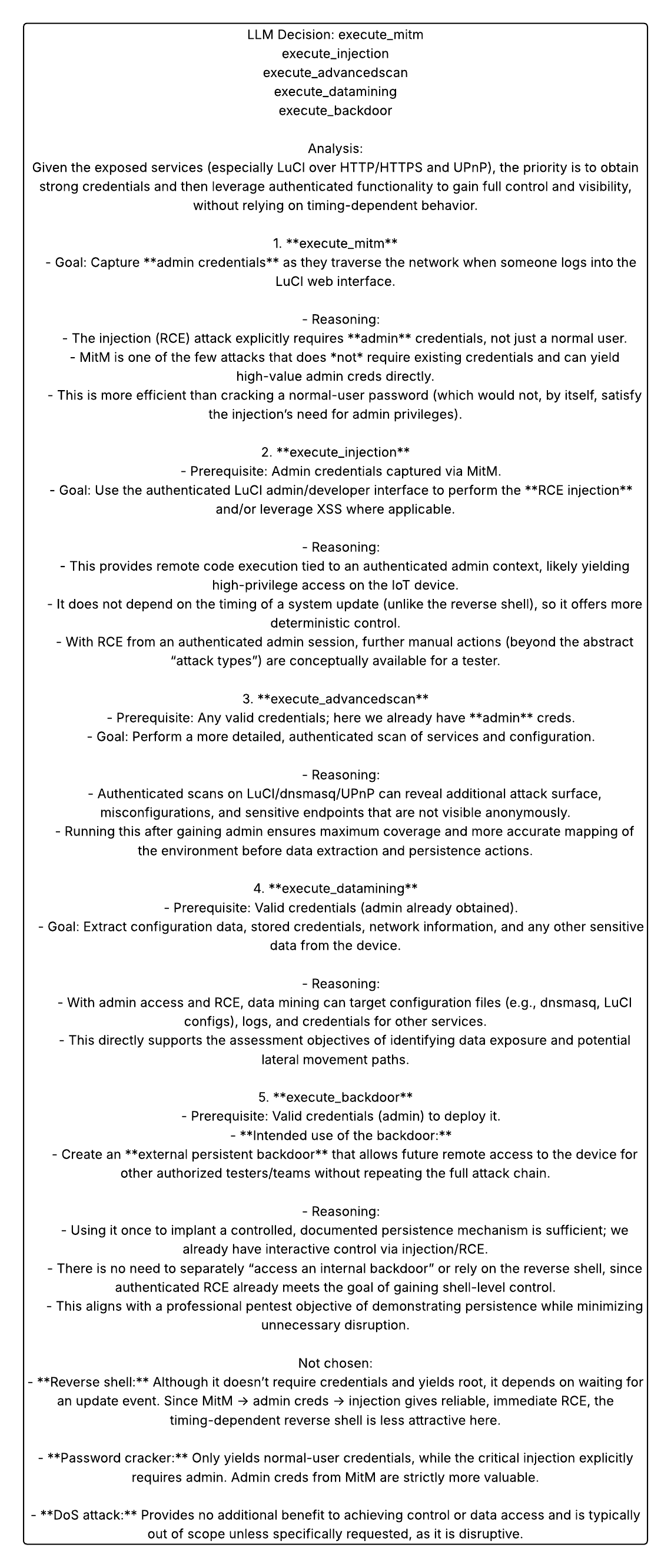}
    \caption{Example output from the vulnerability detection agent showing selected attacks, dependencies, and reasoning for the generated attack sequence.}
    \label{fig:DetectionOutput}
\end{figure}

% The vulnerability detection agent serves as the orchestration component of \reiot. As shown in Algorithm~\ref{Alg:DetectionAgent}, the agent receives a user-defined objective, performs reconnaissance, retrieves vulnerability information, and generates an attack plan using the LLM. The generated plan is parsed to extract attack identifiers, target services, attack parameters, and dependency information.

%\caption{Detection Agent Output}

The vulnerability detection agent serves as the primary orchestration component of the framework. The workflow as presented in Algorithm \ref{Alg:DetectionAgent} begins by scanning the target IoT environment using \texttt{nmap} to identify active ports and running services. The target IP address is provided by the user and corresponds to the vulnerable IoT device deployed within the local testbed environment \cite{nmap}. Each network protocol from the \texttt{nmap} scan is sent individually to the \texttt{searchsploit} tool. It is made up of public exploit scripts for the known CVEs for the given protocol and its specific version. The information for each protocol is stored in a list variable that is inserted into the prompt as well. After these two tools are executed, the prompt is constructed for the API call to the model. 

%The vulnerability detection agent serves as the primary orchestration component of the framework. The workflow as presented in Algorithm \ref{Alg:DetectionAgent} begins by scanning the target IoT environment using \texttt{nmap} to identify active ports and running services. The target IP address is provided by the user and corresponds to the vulnerable IoT device deployed within the local testbed environment \cite{nmap}.
% The results of the scan are stored in a variable that is used during the creation of the prompt. After this reconnaissance is completed, the \texttt{searchsploit} tool is used. Each network protocol from the \texttt{nmap} scan is sent individually to the \texttt{searchsploit} tool. The \texttt{searchsploit} has it's own Exploit Database that it looks through \cite{searchsploit}. It is made up of public exploit scripts for the known CVEs for the given protocol and its specific version. The information for each protocol is stored in a list variable that is inserted into the prompt as well. After these two tools are executed, the prompt is constructed for the API call to the model. 

\begin{algorithm}
\caption{Workflow for Detection Agent}
\begin{algorithmic}[1]
    \STATE \textbf{Input} \textsc{User-defined attack objectives, target IP address}
            \STATE Scan target device using \texttt{nmap}
            \STATE Query \texttt{searchsploit} for vulnerabilities and exploit scripts
            \STATE Generate prompt using reconnaissance and vulnerability data
            \STATE LLM analyzes vulnerabilities and determines attack sequence
            \STATE Call attack execution agent for selected attack
            \STATE Receive attack execution results
            \IF{Active Attack}
                \STATE Determine if the attack succeeded or failed.
            \ELSIF{Passive Attack}
                 \STATE Determine if the attack gathered the information that was needed. If not, mark as a fail.
            \ENDIF
            \IF{Succeeded}
                \STATE Call attack agent for next attack in the sequence.
            \ELSIF{Failed}
                 \STATE Retry attack with modified execution strategy
            \ENDIF
    \STATE \textbf{Output} \textsc{Validated attack sequence and execution results}
\end{algorithmic}
\label{Alg:DetectionAgent}
\end{algorithm}

The LLM analyzes the input to determine what exploits can be used against the environment. It uses it's own knowledge with the results from \texttt{searchsploit's} Exploit Database \cite{searchsploit} to determine an attack that can be used. Some attacks require credentials and information gathered during earlier attack stages. 
Six of the attacks need a password to succeed and two of these attacks require elevated user credentials.  An example of the detection agent's output can be seen in Figure \ref{fig:DetectionOutput}. It shows how each attack is listed as part of the plan with an analysis that justifies each choice and what goal it achieves.

%The attack and the relevant information are extracted and sent to the attack agent. Results are returned from the attack agent. This allows the agent to adjust the approach for success.

The outputs generated by the detection agent are parsed to extract attack identifiers, attack parameters, and execution instructions before being passed to the attack execution agent. If there are multiple attacks, agents sets up a loop through each individual attacks. For an active attack, error or failure messages are collected and sent to the attack agent. For a passive attack, it confirms that the information needed for the attack was collected. If the information was not collected the attack will be executed again. Any failures or errors are handled the same as the active attack. If the attack succeeds the next attack in the sequence is sent for execution. For attacks with dependencies to prior attacks, the information is stored and updated locally on the attacker machine for the vulnerability detection agent to use while calling the attack agent for each attack execution.

\reiot~also supports parallel execution of attacks when dependencies do not conflict.
This is done by changing the code to call multiple instances of the attack agent at once instead of one by one. Attacks with dependencies, like the need for credentials, will wait for the results of that attack before being called, but all attacks that require that dependency will also run at the same time. 

%There is also a way to run the attacks in parallel. The validation for success and failure does not change in this scenario.

%, including vulnerability details, target protocols, exploit descriptions, and attack objectives

% The attack execution agent receives a structured attack request from the vulnerability detection agent. This request includes the attack type, target service, port, vulnerability description, candidate exploit information, and any required prerequisites. As shown in Algorithm~\ref{Alg:AttackAgent}, the agent identifies an appropriate tool or script, generates the execution command, runs the command in the attacker environment, and captures the resulting output.

% The attack execution agent returns a structured status message indicating whether the attack succeeded, failed, or requires retry. It also returns relevant execution details, such as captured credentials, command output, error messages, or observed target-side effects. These outputs are then used by the vulnerability detection agent to update the attack state and determine the next action.

%This agent interacts with all the scripts and tools involved with attack execution.

\subsection{Attack Execution Agent}
\begin{figure}[h]
    \centering
    \includegraphics[width=\linewidth]{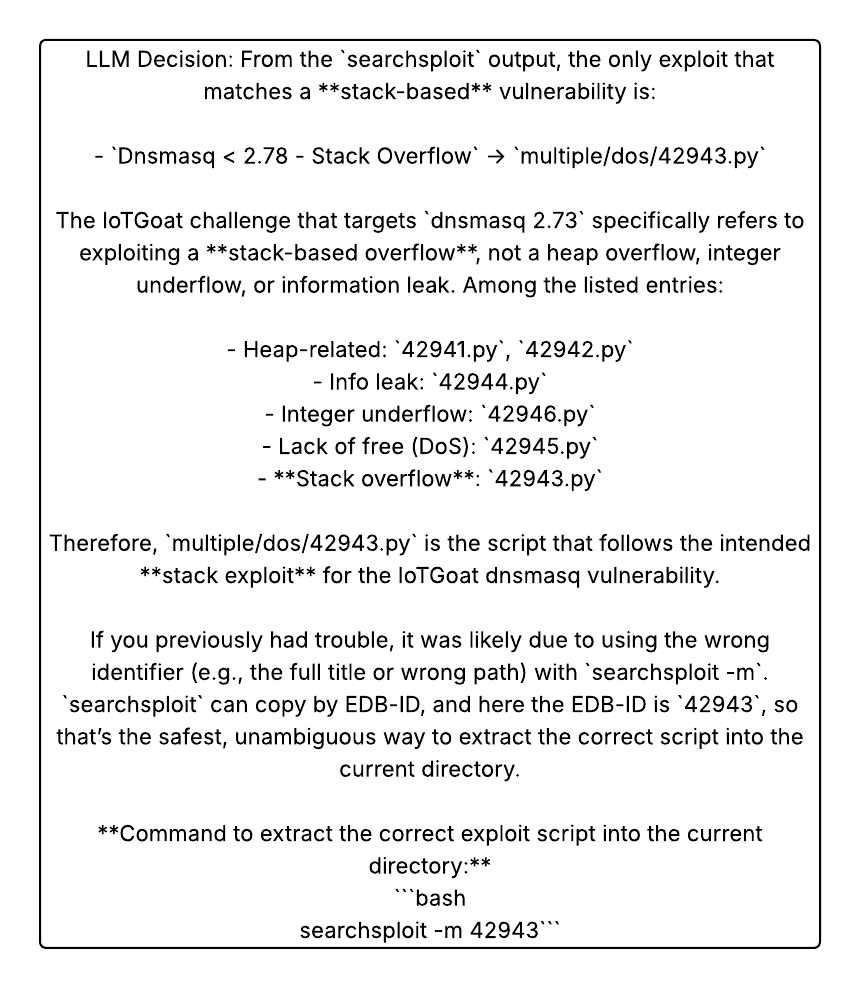}
    \caption{Example output from the attack execution agent selecting a DNS DoS exploit script and generating the corresponding command.}
    \label{fig:AttackAgentOutput}
\end{figure}

The attack execution AI agent is responsible for selecting attack-specific tools, gathering executable commands, and validating attack outcomes. The agent receives structured attack information from the vulnerability detection agent. The workflow for attack execution agent is given in Algorithm \ref{Alg:AttackAgent}.
The attack prompt used by attack execution agent is generic, and the information for a specific attack is inserted from the plan made by the detection agent. Scripts come from the \texttt{searchsploit} tool and stored in the directory corresponding to the attack. An example of how this is stored is seen in Figure \ref{fig:AttackAgentOutput}. The decision is indicated and justified similarly to each attack in the detection agent's output. The command for getting the script and running it is given in bash at the end so the agent can run it. The model also uses its own information to examine the protocol and confirm the presented vulnerabilities and exploits. 

%The general process the agent goes through can be seen in the algorithm below. 

%\hl{I am throwing ideas here: Have discussions on topics aroundL Planning loop, failure handling, termination conditions, output validation, how are attacks selected or ordered, Human in the loop, pseudocode for control loop, prompt templates in an appendix, success validation logics for each attacks, model configuration details if any, false positives in vunerability identifications, some ablation study (What if without stored knowledge, script in out case), }

\begin{algorithm}
\caption{Workflow for Attack Execution Agent}
\begin{algorithmic}[1]
    \STATE \textbf{Input} \textsc{Attack type, vulnerability context, prerequisites from detection agent}
            \STATE Parse attack instructions and exploit details
            \STATE Identify relevant tools or exploit scripts
            \STATE Generate attack execution commands using the LLM
            \STATE Execute generated commands in attacker environment
            \STATE Capture attack outputs and execution results
            \STATE Attack success verification
    \STATE \textbf{Output} \textsc{Attack status and execution results}
\end{algorithmic}
\label{Alg:AttackAgent}
\end{algorithm}

% \begin{algorithm}
% \caption{Attack Agent}
% \begin{algorithmic}[1]
%     \STATE \textbf{Input} \textsc{Attack from detection agent that is being executed and vulnerability information.}
%             \STATE Parse for the specific attack indication in the output.
%             \STATE Extract file path for attack specific scripts.
%             \STATE Determine which tool could be used for the type of attack indicated.
%             \STATE $LLM Model \gets 5$
%             \STATE Model assesses the goal of the attack from the detection output's details.
%             \STATE Model confirms vulnerabilities and exploits with its own knowledge.
%             \STATE Model compares the goal of the attack and identified vulnerabilities for similarity.
%             \STATE Model compares tool and script functionalities. 
%             \STATE Model finds the tool that matches the goal of the attack and indicates this tools should be used.
%             \STATE The command to execute the attack is given.
%     \STATE \textbf{Output} \textsc{Commands for executing the attack.}
% \end{algorithmic}
% \label{Alg:AttackAgent}
% \end{algorithm}

%\lopa{Is there any logic how llm select thhe attack? that means what is the input to llm model}

%\caption{Attack Agent Output}

The result of this reasoning for the DoS attack can be seen in Figure \ref{fig:AttackAgentOutput}, where the command in the output is executed to use the specific tool or script. After the execution of the attack, the results of the attack are received. For passive attacks, information is returned while active attacks show the impact on the target device. The results of attack execution are returned to the vulnerability detection agent, which validates the outcome and determines whether additional retries or subsequent attacks should be executed.

\subsection{Attack Validation}

% The validation for each of the 10 attacks varies due to the diversity of the attacks that are included. Passive attacks are simple and just wait for the information to be received. The MitM attack succeeds if the admin username and password is intercepted. Active attacks have to confirm that certain actions were completed creating more complexity. The XSS, RCE, and update attacks receive a confirmation that a payload was injected by probing what was edited and confirming it is what was expected. Other attacks like cracking a hard-coded password and gathering database information are validated by ensuring the information was gathered similar to passive attacks. The log erase attack simply checks to see if any logs still exist after running the attack. The DoS attack uses a buffer overflow message and uses the confirmation that the packet was received as validation. The attack that opens a backdoor is confirmed by a success message and accessing that backdoor. The developer backdoor has a successfully connected message that validates the attack's success. These validation processes will also return errors, unexpected payloads, or lack the expected information which can be used to determine a failure. 

Because the evaluated attacks differ in objective and execution behavior, \reiot~uses attack-specific success criteria. Passive attacks are validated by confirming that the expected information was collected, while active attacks are validated by observing command execution, service disruption, file modification, or successful access to a target service. Table~\ref{tab:validation} summarizes the validation criteria used for each attack.

\begin{table}[t]
\centering
\caption{Attack Validation Criteria}
\label{tab:validation}
\renewcommand{\arraystretch}{1.2}{
\begin{tabular}{p{0.28\linewidth}p{0.09\linewidth}p{0.44\linewidth}}
\hline
\textbf{Attack} & \textbf{Type} & \textbf{Success Criterion} \\
\hline
MitM credential grab & Passive & Administrative credentials are intercepted from client-gateway traffic. \\
Hardcoded password & Active & Credentials are recovered and can be used for authentication. \\
MiniUPnP backdoor & Active & Backdoor access is created and reachable. \\
Developer backdoor & Active & Connection to the developer backdoor succeeds. \\
XSS & Active & Injected payload executes or produces the expected webpage effect. \\
RCE & Active & Injected command produces the expected file or system modification. \\
Update attack & Active & Malicious update payload executes successfully. \\
DNS DoS & Active & DNS service becomes unavailable and remains down until restart. \\
Database PII & Active & Targeted sensitive records are retrieved from the database. \\
Log erase & Active & Relevant logs are removed from the target system. \\
\hline
\end{tabular}}
\end{table}

\section{Implementation and Results}
\label{sec:implementation}
The proposed \reiot~framework is evaluated in controlled vulnerable environments to assess its ability to autonomously identify and exploit IoT-related vulnerabilities. Although the framework was designed to remain adaptable across heterogeneous IoT systems, IoTGoat \cite{IoTGoat} was selected as the primary evaluation environment because it provides intentionally vulnerable firmware implementations mapped to OWASP IoT Top 10 \cite{OWASP}. A dedicated testbed consisting of attacker, client, and target virtual machines was constructed to evaluate attack execution and attack dependencies. Additional experiments were also conducted in a Metasploitable2 \cite{Rapid7} environments to evaluate the framework beyond IoTGoat specific vulnerabilities.

% While this framework was made to be agnostic for different IoT devices in order to achieve flexibility when testing IoT vulnerabilities, a specific environment was chosen to showcase several vulnerabilities on one machine that could be exploited for unique attacks. A testbed was set up using this environment and a common attacker machine. The results of the agent in this set up were gathered and showed the effectiveness of the framework.

\subsection{Experimental Setup}

\subsubsection{IoTGoat and Vulnerability Mapping}

IoTGoat implements intentionally vulnerable IoT firmware behaviors corresponding to common OWASP IoT security risks\cite{OWASP}. IoTGoat covers 9 of the OWASP IoT Top 10 categories excluding ``Lack of Physical Hardening'' category as it relates to physical device design rather than firmware or network behavior. The ``Insecure Ecosystem Interfaces'' category contains two distinct exploitable interfaces, resulting in 10 evaluated IoTGoat attack scenarios.

Table~\ref{Table:VulleranbilitMapping} summarizes the vulnerability categories, concrete vulnerabilities, and corresponding attack scenarios used in the evaluation. These scenarios cover credential recovery, insecure service exploitation, web-based injection, insecure update behavior, denial-of-service, sensitive data extraction, and log manipulation.

\begin{table*}[!t]
\centering
\caption{OWASP IoT Vulnerability Mapping and Evaluated Attack Scenarios}
\label{Table:VulleranbilitMapping}
\renewcommand{\arraystretch}{1.15}
\begin{tabularx}{\textwidth}{l X X}
\toprule
\textbf{OWASP Category} & \textbf{Implemented Vulnerability} & \textbf{Evaluated Attack Scenario} \\
\midrule
Weak Passwords & Default user password stored in firmware and recoverable by cracking. & Recover default user credentials using \texttt{John the Ripper}. \\ \hline 
Insecure Network Services & MiniUPnP service allows unauthorized port manipulation. & Create a network-accessible backdoor. \\ \hline
Insecure Ecosystem Interfaces & LuCI webpage and developer backdoor are exposed. & Execute XSS and access developer backdoor without credentials. \\ \hline
Lack of Secure Update Mechanism & OpenWrt package update mechanism accepts malformed hash behavior. & Trigger malicious update behavior. \\ \hline
Use of Outdated Components & \texttt{dnsmasq} 2.73 contains known DoS vulnerabilities. & Execute DNS denial-of-service attack. \\ \hline
Insufficient Privacy Protections & Database stores sensitive user information in plaintext. & Extract PII from the database. \\ \hline
Insecure Data Transfer & HTTP is used for sensitive web interactions. & Intercept traffic and administrative credentials using MitM. \\ \hline
Lack of Device Management & Logs are stored locally without external management. & Remove local logs after attack execution. \\ \hline
Insecure Default Settings & Web interface allows CSRF-style command execution. & Use forged web interaction for RCE. \\ 
\bottomrule
\end{tabularx}

\end{table*}

\subsubsection{Testbed}

\begin{figure}[!t]
    \centering
    \includegraphics[width=\linewidth]{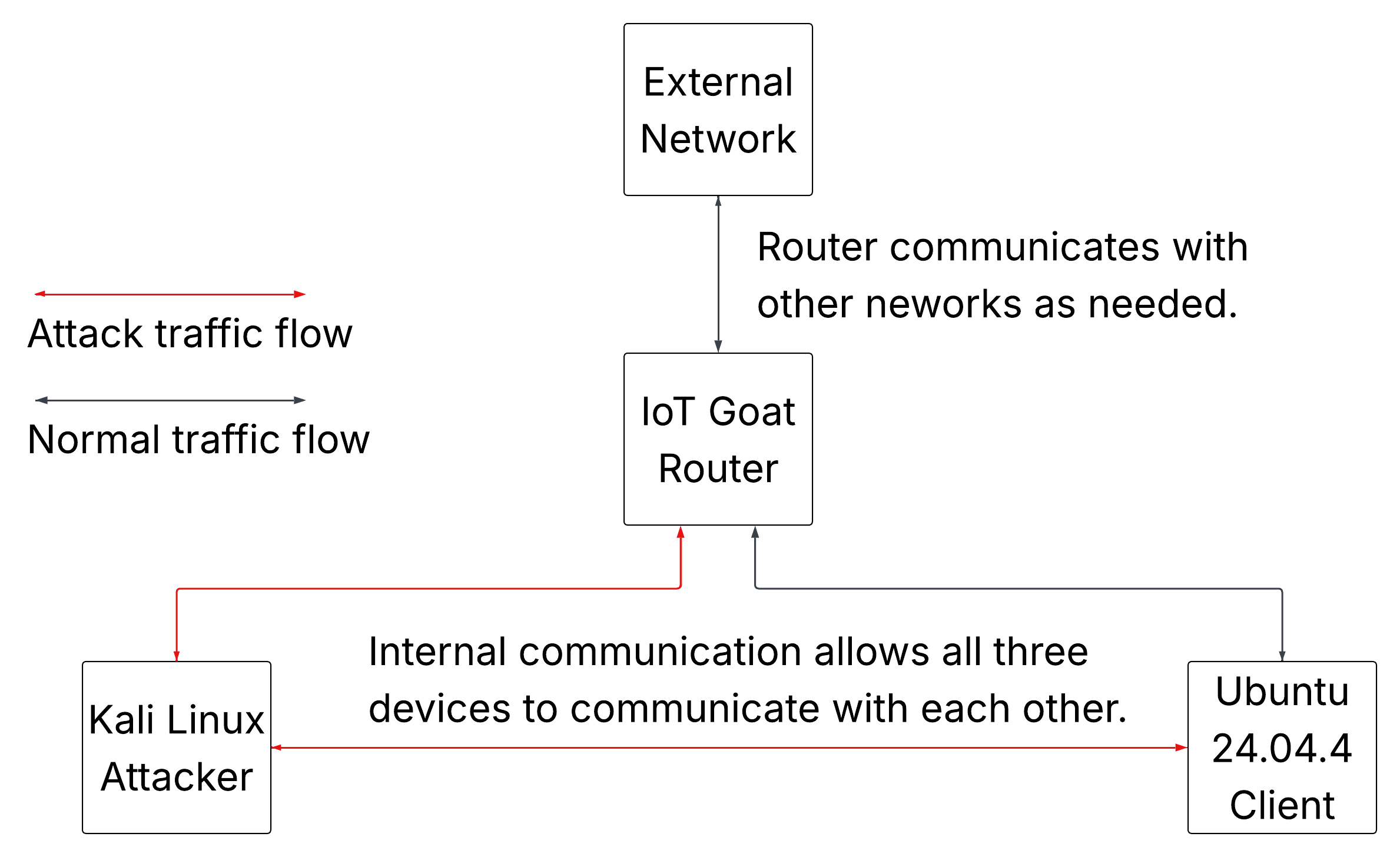}
    \caption{IoTGoat Testbed}
    \label{fig:goat_testbed}
\end{figure}
Figure \ref{fig:goat_testbed} shows the complete testbed configuration and includes details about additional tools that were downloaded and specific OS systems that are running.
The testbed for this framework uses 3 virtual machines in a attacker, client, and server configuration. The IoTGoat virtual machine serves as the target device and was configures as a router running \texttt{OpenWrt} firmware. 
The particular network configurations came from the suggested setup found on the IoTGoat GitHub \cite{IoTGoat}. Kali Linux was used as the attacker machine and hosted the proposed agents along with offensive security tools including \texttt{nmap}, \texttt{searchsploit}, \texttt{bettercap}, and \texttt{John the Ripper}. An Ubuntu 24.04.4 virtual machine acted as the client device to support man-in-the-middle (MitM), DNS denial-of-service, and cross-device XSS evaluation scenarios.

% The client is an Ubuntu 24.04.4 virtual machine. The client was used to perform MitM attacks and helped show the effects of the DoS and XSS attack. The most recent version of Kali Linux was downloaded from the official website. The Kali Linux machine houses the attack agents and uses the pre-installed tools and bettercap. 

%Attacks can be launched on the IoTGoat machine, but also has malicious communications with the client machine. The testbed was setup this way to ensure every attack could be implemented. The client machine was necessary for the MitM attack to be performed. The impact from DoS attack on the DNS service and the XSS attack were also observable from the client.  The DoS on the DNS is seen on the client because it cannot access any websites without the ip address when the service is down. The XSS attack shows the firewall rule that was injected and is visible on another machine.
The experiments assume an attacker with network-level access to the target IoT environment and access to common offensive security tools available on Kali Linux. Physical access to the target device is not assumed. All experiments were conducted in isolated virtualized environments to ensure controlled and reproducible testing conditions.

\subsection{Experimental Results}
The framework was evaluated using attack success rate, execution time, LLM reasoning time, and token usage metrics. Each attack scenario was executed 20 times to evaluate execution consistency, failure behavior, and model reliability. Success was determined based on attack-specific validation criteria, including credential interception, payload execution, service disruption, and successful exploit completion. The LLM model used for these experiments was ChatGPT 5.1 thinking.

% \textbf{Attack Success Rate:} Table \ref{tab:success} presents the success rates for the 10 attacks executed by the attack agent, along with the number of successful trials for each attack. In all 200 trials, the vulnerability scanning agent operated without failure, resulting in a 100\% success rate. 7 of the 10 attacks also had a 100\% success rate and the lowest success rate was 80\%.

\textbf{Attack Success Rate:} Table~\ref{tab:success} presents the attack execution success rates of the \reiot~framework across the 10 IoTGoat attack scenarios. Each attack was executed 20 times, resulting in 200 total attack execution trials. The vulnerability detection agent successfully completed reconnaissance and vulnerability planning in all 200 trials, indicating that failures occurred during attack execution rather than during vulnerability discovery or attack planning. Overall, the attack execution agent successfully completed 189 out of 200 attack trials, corresponding to a 94.5\% overall success rate. Seven of the 10 attacks achieved a 100\% success rate, including XSS, developer backdoor access, update attack, database PII extraction, log erase, and RCE. These results suggest that the framework is highly effective when the attack workflow has clear execution steps, well-defined validation criteria, or direct tool/script mappings. The lowest success rates were observed for the MiniUPnP backdoor and DNS DoS attacks, each achieving 80\% success. These attacks required more precise command construction or interaction with service-specific behavior, making them more sensitive to syntax errors and model refusals. The MitM credential grab achieved 95\% success, showing that passive interception workflows were generally reliable once the required network conditions were established. These results indicate that \reiot~is most reliable for attacks with deterministic execution paths and explicit success conditions, while attacks requiring highly parameterized commands, service-specific exploit behavior, or refusal-prone instructions remain more challenging.

\begin{table}{}
\caption{Attack Success Rates of \reiot~Framework in IoTGoat Environment}
\label{tab:success}
{\renewcommand{\arraystretch}{1.1}
\begin{tabular}{|c|c|c|c|} 
\hline
\textbf{Attack} & \textbf{Success} & \textbf{Failed} & \textbf{Total} \\ \hline \hline

Hardcoded user password & 18 & 2 & 20 \\ \hline

Open MiniUpnp backdoor & 16 & 4 & 20 \\ \hline

XSS Attack & 20 & 0 & 20 \\ \hline

Developer Backdoor & 20 & 0 & 20 \\ \hline

Update Attack & 20 & 0 & 20 \\ \hline

DNS DoS Attack & 16 & 4 & 20 \\ \hline

Database PII & 20 & 0 & 20 \\ \hline

MitM credential grab & 19 & 1 & 20 \\ \hline

Log erase & 20 & 0 & 20 \\ \hline

RCE Attack & 20 & 0 & 20 \\ \hline
\end{tabular}}
\end{table}

\textbf{Failure Analysis:} 
Three primary causes of attack failure were observed during experimentation: model refusals, persistent syntax-generation errors, and hallucinated outputs. Table \ref{tab:fails} shows reasons of failure and its frequency across all the attacks by attack execution agent. Model refusals occurred when the LLM declined to generate attack-related instructions. These accounted for two MiniUPnP failures and three DNS DoS failures. Syntax-generation failures occurred when generated commands contained invalid or incomplete parameters that were not corrected during retry attempts. These occurred twice in the hardcoded password attack, twice in the MiniUPnP attack, and once in the DNS DoS attack. Hallucination-related failures were less common and occurred only once during the MitM credential interception experiments. Most syntax-related failures occurred during attacks requiring longer or highly parameterized command generation, suggesting that complex command composition remains a challenge for autonomous LLM-driven execution workflows.

\begin{table}{}
\centering
\caption{Failure Causes in IoTGoat Experiments} 
\label{tab:fails}
{\renewcommand{\arraystretch}{1.1}
\begin{tabular}{|l|c|} % Replaced X with l (left-aligned)
\hline
\textbf{Reason for Failure} & \textbf{Frequency in Trials} \\ \hline 
Hallucination & 1 \\ \hline
Persistent Syntax Error & 5 \\ \hline
Model Refusal & 5 \\ \hline
\end{tabular}}
\end{table}

% There were 3 reasons for the failures seen in the trials. Table \ref{tab:fails} shows each reason and how often it occurred in the trials. One reason was the model refusing to give the information that could be used for the attack. This happened 2 times with the MiniUpnp attack and 3 times with the DoS attacks. The second reason was due to persistent syntax errors that were not fixed by the agent which made up other failures with the DoS and MiniUpnp attacks and the 2 failures with the hardcoded password. The last reason for failure was hallucinations. This only happened once with the MitM attack.

\textbf{Execution Time and Token Usage:} 
Execution time, LLM reasoning time, and token usage were recorded for each successful attack execution. Since attack workflows varied significantly in complexity and required contextual input, token usage and reasoning time differed across attack categories. These results can be used to compare efficiency of the agent compared to traditional adversarial methods and to help determine possibilities for further scaling. Table \ref{tab:metrics} shows the results for each attack the agent ran averaged over the 20 trials. Most of the attacks were able to be executed in less than a minute while the longest attack was under 2 and a half minutes. For most attacks, the LLM reasoning took the majority of the time. The exception to this is the hardcoded user password because it takes a longer time to crack a password. The time the LLM reasoning took was less than 30 seconds for most attacks that had simple port numbers or webpage access as an input and used between 500 to 700 tokens. Because some attacks like log erase and the database PII attack required looking through files containing logs and directories, they took 45 to 50 seconds for reasoning. This affected both the time taken for reasoning and the token usage so attacks with more input and parsing used twice as many tokens and took twice the amount of time.

\begin{table}[t]
\centering
\caption{Execution Time and Token Usage for \reiot~Framework in IoTGoat}
\label{tab:metrics}
\renewcommand{\arraystretch}{1.2}
% Wrap the tabular environment in a resizebox
\resizebox{\columnwidth}{!}{%
\begin{tabular}{|l|l|l|l|} 
\hline
\textbf{Attack} & 
\textbf{\begin{tabular}[c]{@{}l@{}}Mean Time \\ (seconds)\end{tabular}} & 
\textbf{\begin{tabular}[c]{@{}l@{}}Mean Time for \\ LLM Reasoning\end{tabular}} & 
\textbf{Mean Tokens} \\ \hline \hline

Hardcoded user password & 139 & 28 & 690 \\ \hline
Open MiniUpnp backdoor & 28.44 & 26.15 & 530 \\ \hline
XSS Attack & 24.53 & 19.97 & 670 \\ \hline
Developer Backdoor & 46.03 & 44.04 & 520  \\ \hline
Update Attack & 51.12 & 49.92 & 540 \\ \hline
DNS DoS Attack & 37.81 & 22.32 & 920 \\ \hline
Database PII & 102 & 54.39 & 1480 \\ \hline
MitM credential grab & 15.63 & 14.44 & 620 \\ \hline
Log erase & 54.00 & 53.51 & 1390 \\ \hline
RCE Attack & 13.03 & 11.75 & 460 \\ \hline
\end{tabular}%
} % End resizebox
\end{table}

\textbf{Observed Attack Outcomes:} 
The \textit{hardcoded password} attack compares a list of known passwords and the hashes to what is in the firmware to get a username and password. These are user credentials not admin, but it still provides access that can allow commands to be run and a shell to be opened. Every time the password cracking started, the username and password was successfully found. The \textit{MiniUPnP attack} uses the service to open a new ssh port that can be used by the attacker. Because it is native to the service, it is successful as long as the command is given and a new reachable service path was created and validated after successful execution.

The \textit{XSS attack} added a new firewall rule through the developer webpage. To show impact from it, the background of the webpage turned red when the new rule was clicked on. For each trial, the webpage was accessed to confirm this worked before deleting the rule and trying again. The \textit{update attack} waits for an update to be triggered and intercepts it with a command to open a shell on an open port on the attacker. This is similar to the \textit{developer backdoor} that connects to the port left open by the developers for access. Both of these attacks once given access can run commands input by the user. The backdoors were configured to assume the user wants access to perform actions themselves. 

For the \textit{DNS DoS attack}, the attack script is ran until it shuts down the service and it has to be restarted. The attack takes down the service once it starts and averaged 7 seconds to become permanent and 15 seconds at the longest. The availability for the service stays down after the attack as well and only comes back up if it is restarted leading to 0\% availability during the time after the attack and before the service is restarted. The IoTGoat database had sensor data tied to fake users. The \textit{database PII}  attack simply pulls the information and prints it showing how each user interacts with the sensor including sensitive information like location coordinates. 

The \textit{MitM attack} was passive interception. Success for this is measured with how many packets were intercepted between the client and gateway victim. For all trials, 100\% of the packets were intercepted showing the MitM attack was successful at intercepting traffic data. It also got 100\% of the sensitive data, which in this case was just root username and password, for all the success trials. 

The \textit{log erase attack} was a simple attack that used the lack of log management to clear all the dmesg logs on the machine. This will have logged any suspicious activity like creating a new port with MiniUPnP and the DNS service going down. This was ran in between trials to ensure there were logs to clear and confirmed by checking the dmesg file after. 

 The \textit{RCE attack} worked in a similar way. The rc.local file in the etc directory is wiped and replaced with a command to create directory. This causes the directory to be made when the VM restarts. Each trial included looking for these changes in the file and on restart.

%\hl{Captions for the figures should be descriptive which is not the case for both Fig 7 and 8}

\begin{figure}[h]
    \centering
    \includegraphics[width=\linewidth]{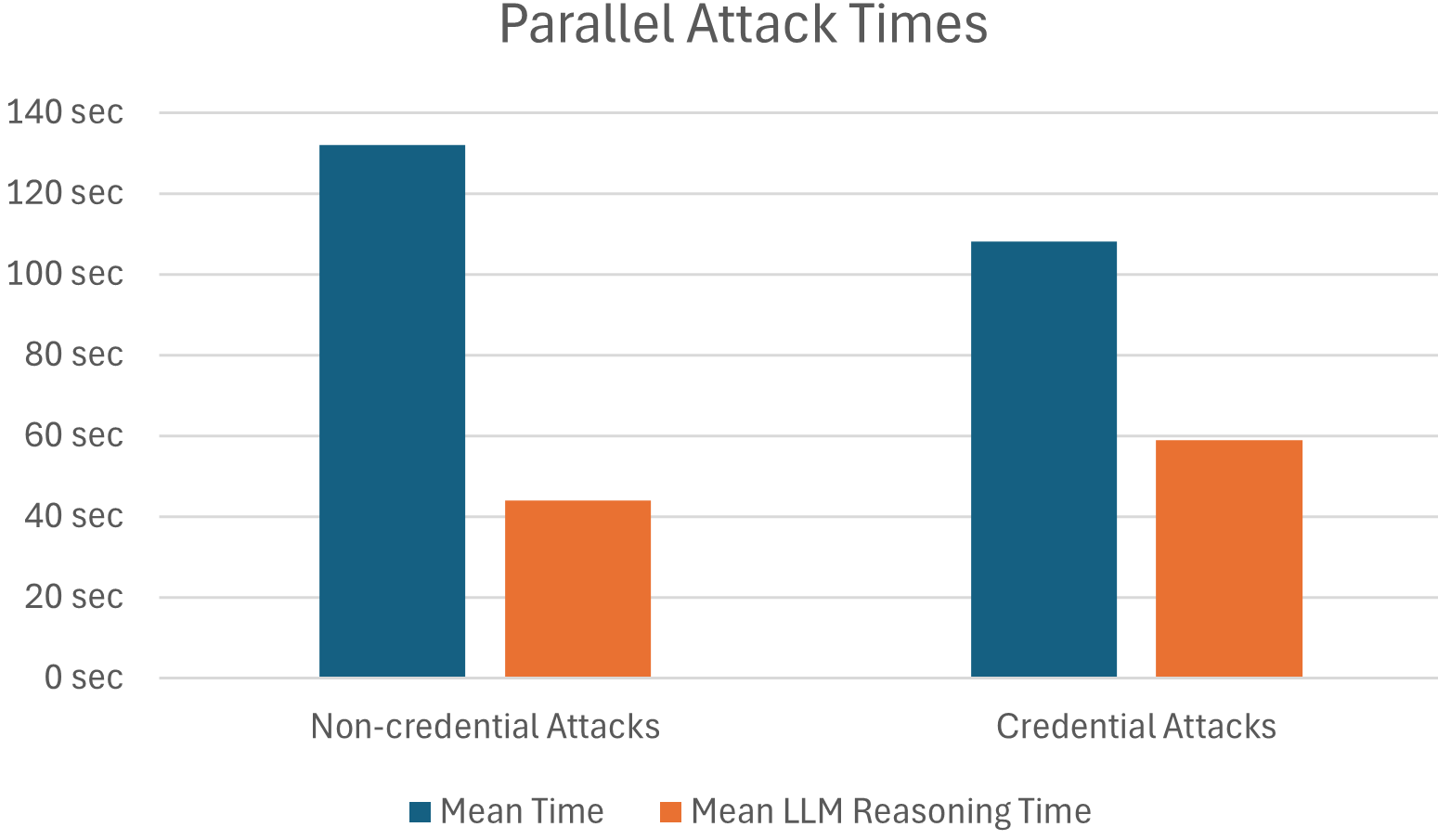}
    \caption{Time taken to run parallel attacks in IoTGoat}
    \label{fig:time}
\end{figure}

\begin{figure}[h]
    \centering 
    \includegraphics[width=\linewidth]{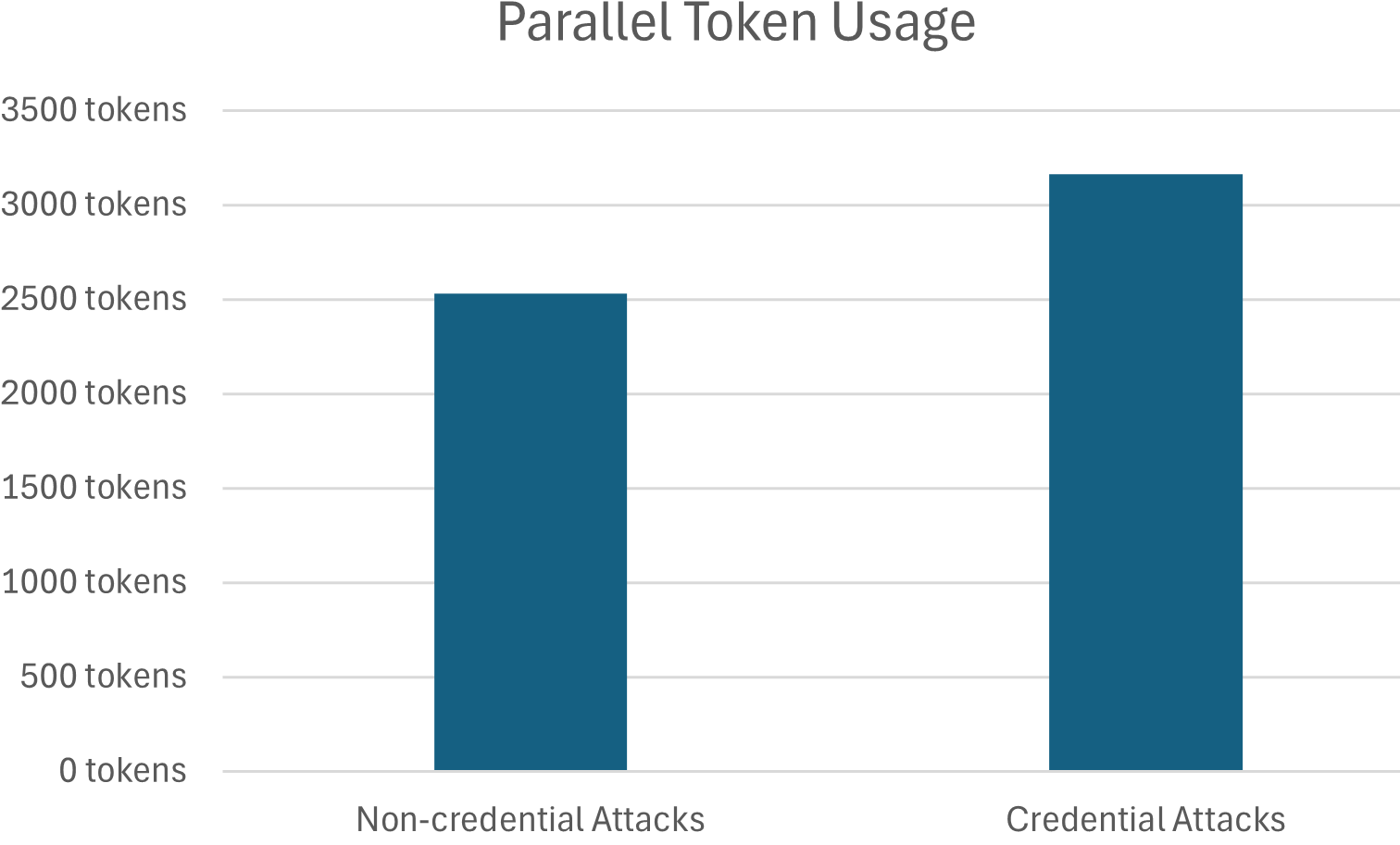}
    \caption{Tokens used to run parallel attacks in IoTGoat}
    \label{fig:tokens}
\end{figure}

\textbf{Parallel Execution:}

The \reiot~framework also supports dependency-aware parallel execution when attacks do not have conflicting prerequisites. To evaluate this capability, the attacks were grouped into two categories: non-credential-dependent attacks and credential-dependent attacks. Non-credential-dependent attacks, such as the DNS DoS attack and developer backdoor access, can be executed without previously collected credentials. These attacks were executed concurrently with credential-gathering attacks, including MitM credential interception and hardcoded password cracking. 
Once credential collection completed successfully, the remaining credential-dependent attacks were executed in parallel. Figures~\ref{fig:time} and~\ref{fig:tokens} show the execution time and token usage for these parallel attack groups. The results show that parallel execution significantly reduces total wall-clock time because the completion time of each group is dominated by the longest-running attack rather than the sum of all attacks in that group. For the non-credential-dependent group, the hardcoded password attack dominated execution time, averaging approximately 2 minutes and 15 seconds. In contrast, the credential-dependent group completed in approximately 1 minute and 35 seconds. As a result, all 10 IoTGoat attacks completed in approximately 3 minutes and 50 seconds under dependency-aware parallel execution while individual execution of all 10 attacks was completed in approximately 8 minutes and 31 seconds.

The token usage results in Figure~\ref{fig:tokens} show a different trend. Unlike execution time, token usage remains approximately additive because each parallel attack still requires a separate invocation of the attack execution agent. The non-credential-dependent group used approximately 2,500 tokens, while the credential-dependent group used approximately 3,100 tokens. The higher token usage for the credential-dependent group is expected because it includes more attacks and several attacks with larger contextual requirements. These results suggest that parallelization improves wall-clock efficiency but does not substantially reduce total LLM token cost. Therefore, \reiot~is scalable in terms of execution time, but token usage remains dependent on the number and complexity of attacks executed.

%\Gupta{need restructuring}

% The attacks could also be ran in parallel. The DoS attack and developer backdoor do not need credentials so these two attacks could be run with the MitM and password cracking attack. These are the non-credential attacks of the 10. After these attacks are ran, the other 6 attacks can be ran in parallel after these attacks. These are the credential attacks. Figure \ref{fig:time} and \ref{fig:tokens} shows the metric results for these attacks.

\subsection{Metasploitable Implementation and Results}

To evaluate the framework beyond IoTGoat-specific vulnerabilities, additional experiments were conducted using the Metasploitable2 environment \cite{Rapid7}. Although Metasploitable2 is not specifically designed for IoT research, it contains several vulnerable services commonly observed in IoT-related systems and has been widely used in cybersecurity experimentation and dataset generation \cite{KORONIOTIS2019779,botnet}. Therefore, it was used as a secondary vulnerable environment to evaluate whether the framework could adapt its reconnaissance, vulnerability lookup, and exploit execution workflow beyond a single target configuration.The testbed setup consisted of 2 virtual machines. Kali Linux remained the attacker virtual machine and the Metasploitable2 virtual machine was the target IoT device.

While Metasploitable2 is a good IoT surrogate, several protocols and their vulnerabilities do not apply to IoT devices and their common vulnerabilities. Due to this, only some vulnerabilities were targets for attacks using our \reiot~framework. Three vulnerable services were evaluated: a backdoor in the code of the vsftdd v2.3.4, the lack of encryption in the MySQL database, and an RCE attack with an insecure distcdd \cite{Rapid7}. The agents then planned and ran the attacks in the new environment and saw similar success to the IoTGoat experiments. Table \ref{tab:tokens} shows the results of these attacks.

\vspace{1em}
\noindent
\begin{minipage}{\linewidth}
\centering 
\captionof{table}{Metasploitable Attack Results} 
\label{tab:tokens}
% This line forces the table to be exactly the width of the column
\resizebox{\linewidth}{!}{%
{\renewcommand{\arraystretch}{1.1}
\begin{tabular}{|l|c|c|c|c|} 
\hline
\textbf{Attack} & \textbf{Success} & \textbf{Failures} & \textbf{Total} & \textbf{Mean Tokens} \\ \hline 
Hardcoded Backdoor & 20 & 0 & 20 & 580  \\ \hline
Database Passwords & 20 & 0 & 20 & 620  \\ \hline
RCE Attack & 18 & 2 & 20 & 520  \\ \hline
\end{tabular}}%
}
\end{minipage}
\vspace{1em}

The backdoor and database attacks were successful in all trials. The 2 failures for the RCE attack came from model refusals. The token usage was close to the counterparts in IoTGoat for the backdoor and RCE attacks. The database was different because a service could be used to access the database instead of searching for it on the machine. This meant the model had less input to process. Compared to other attacks in IoTGoat that used services to execute a command like XSS it also used the expected amount of tokens. The times were also close to the IoTGoat counterparts. The RCE attack averaged 15.04 seconds and 12.09 seconds of that was the LLM reasoning. The backdoor averaged 38.37 seconds and 33.89 seconds for the LLM reasoning. The database attack varied here as well for similar reasons the token usage did. It averaged 22.09 seconds and 18.97 seconds for LLM reasoning which is similar to the XSS attack as well.

\section{Conclusions and Future Work}
\label{sec:conc}
This paper presented \reiot, a dependency-aware multi-agent framework for autonomous IoT vulnerability analysis and exploit execution in controlled testbeds. \reiot~decomposes the workflow into a vulnerability detection agent that performs reconnaissance and attack planning, and an attack execution agent that selects tools, executes attack primitives, and validates outcomes. \reiot~was evaluated across IoTGoat and Metasploitable2 environments. Across 260 total attack executions, the framework achieved a 95.0\% overall success rate. The results demonstrate the feasibility of LLM-driven agents for automating security testing workflows in isolated IoT and IoT-adjacent environments. At the same time, the observed failures show that reliable autonomous exploit execution requires stronger command validation, safety controls, and error recovery mechanisms.

Future work will extend \reiot~with schema-constrained command generation, retrieval-augmented vulnerability reasoning, larger IoT firmware testbeds, human-in-the-loop approval policies, and defensive agents capable of detecting, prioritizing, and mitigating vulnerabilities discovered by adversarial agents.

% This work showed that LLM based AI agents can be used for adversarial purposes in IoT environments. The framework was designed to be both model agnostic and work in different environments. It focused on IoTGoat to target IoT specific vulnerabilities and tested in Metasploitable as well showing how it is adaptable and can be integrated into existing workflows. The results show that this is a viable, effective way to quickly exploit vulnerabilities in multiple IoT environments. There is also the potential to add a human-in-the-loop implementation. The agents are designed to be completely autonomous to see the full potential for efficiency and speed, but is also set up to be monitored by a human by printing everything. The design could be altered slightly to need human approval of the plan and results before the agent moves forward. It could also allow user alterations to help fix the persistent syntax errors seen in the results section. Broadly, the \reiot can be used for large scale security vulnerability scanning and exploitation in IoT environments. 

% The next step is to use agents to mitigate the attacks launched by the adversarial agent. Ideally, it will be setup similarly to the blue team counterpart with an agent for detecting the vulnerabilities on-device and a general agent that executes mitigations based on the information it is given. It would also be interesting to have a proactive approach to patching problems when possible while still having real-time responses to attacks. 

%\iffalse
\section*{Acknowledgment}

This work is partially supported and developed under the National Science Foundation Award \# 2346001 NRT-GCR, AI: Immersive Research Traineeship in the Convergence of Artificial Intelligence, Energy and Cyber Security.

%\fi
\bibliographystyle{IEEEtran}
\bibliography{references}
\begin{comment}

\vspace{12pt}
\end{comment}
\end{document}